# Infrared Passbands for Precise Photometry of Variable Stars by Amateur and Professional Astronomers

**Eugene F. Milone**
*Department of Physics and Astronomy, University of Calgary, 2500 University Drive, NW, Calgary, AB T2N 1N4, Canada*

**Andrew T. Young**
*Astronomy Department, San Diego State University, PA-210 5500 Campanile Drive, San Diego, CA 92182-1221*

**Abstract**   The infrared (IR) spectral region is a rich one for variable star work, especially of cooler stars, but it is hard to do IR photometry because of high, variable background, and specialized telescopic equipment that is usually required to obtain meaningful data. Typically, telescopes with IR detectors are at high elevations, to minimize water vapor absorption. Nearly all the filters produced for astronomical work at observatories around the world have not been optimized for use at anything other than the highest and driest of observatories. This has made it difficult for amateur astronomers to contribute to this field. Now, however, this is no longer the case. The IAU's Infrared Working Group (IRWG) has designed and tested a set of IR filters less sensitive to water vapor, permitting observations at any site where precise optical photometry can be carried out. Data acquired with these filters can be corrected easily for atmospheric (water vapor) extinction, unlike the situation with the older IR filters. We demonstrate this with data from the University of Calgary's Rothney Astrophysical Observatory.

## 1. Introduction

Infrared radiation was discovered in the light of the Sun by William Herschel (1738–1822). Herschel (1800) used thermometers to quantify the heat that passed through each of the colors. The highest temperature was found just beyond the reddest color. Despite this promising beginning, it took a while before IR detectors were sufficiently sensitive to permit other objects to be observed. Charles Piazzi Smyth discovered lunar IR radiation in 1856 with a thermocouple at Tenerife. In 1880, Lord Rosse observed the Moon through its phase cycle to determine its temperature.

Bolometers for wavelengths longer than ~2.5 micrometers, and lead sulfide (PbS) detectors with relatively low sensitivity and long time constant in the near IR (~1 to 2.5 micrometers), were used up to the 1960s to observe bright stars and planets. In the 1970s newer detectors became available, and a wider range of objects could be observed. Currently, InSb and HgCdTe detectors and imaging arrays are widely used for infrared detection. One manufacturer has now produced a relatively inexpensive infrared detector for use by amateur astronomers, which we discuss below. Consequently, infrared photometry can be done by any dedicated observer, if the conditions under which it can be carried out are understood and practiced carefully. We will discuss these shortly. First, however, we consider why one should observe in the IR and what kinds of objects can be observed optimally.

## 2. Interesting astronomical targets in the infrared

Currently, objects that can be studied include:

- Binary stars
- Eclipsing binaries
- Pulsating stars
- Eruptive stars (novae and SN ejecta)
- Occultations of extrasolar planets
- Interstellar and circumstellar dust
- Earliest galaxies

Because this is not a treatise on the infrared emission characteristics of these objects, for a few cases we give just a brief sketch of why they can be observed usefully in the infrared. Any systematic study of an object should be started only after fully reviewing its properties and potential limitations of photometry to reveal them.

The infrared colors of visible binary components provide important clues to their natures. For the very cool objects known as brown dwarfs, for example, the colors help to identify temperatures and their intrinsic brightnesses indicate ages. Often, IR spectroscopy is called on to refine our knowledge of these and other cool objects.

For eclipsing binaries, the UV is sensitive mainly to blue companions and plage regions, whereas the IR is sensitive to red companions and can help model cool star spots. Broad passbands covering the entire spectral region therefore provide better constraints on the temperatures of both components and high and low temperature regions on both stars than do passbands covering only a small spectral region. Such capability will be essential to study the ground-based light curves and radial velocity curves that will be needed to follow up millions of eclipsing binary discoveries made by existing and planned surveys, such as that from the space-based GAIA mission scheduled to be launched in 2011. Specific examples of the IR advantage can be seen in the *B* and *K* simulated passbands of the Algol system DM Persei, plotted with an earlier version of BINARY MAKER than the currently marketed version, 3.0 (Bradstreet and Steelman 2004) in Figure 1. These data, based on the observed light curves of Hilditch *et al.* (1992) and simulations of Terrell *et al.* (1992), show the relative depth changes of the primary and secondary eclipses (where the hotter and cooler stars are eclipsed, respectively). In passing, we note that a new solution for this system has just been published by Van Hamme and Wilson (2007), who have a different set of parameters, including the temperatures of both components. They found temperatures of 18,000 and 7816 K (giving a ratio of 2.3 in temperature, and of 28 in surface brightness), corresponding to spectral types B2.5 and A1.5 for stars 1 and 2, respectively. Note that the secondary is not a cool star, but it is far cooler than its companion.

Among pulsating stars, amplitudes of Cepheids depend strongly on temperature variation and so are larger in the ultraviolet (UV); in infrared passbands, on the other hand, the amplitudes are smaller. As the temperature changes during a cycle, the flux increase due to this cause is proportionately less in the infrared, so the IR flux is more sensitive to size variation. Photometry is always welcome, if even over a single cycle and in a limited number of passbands (but two or three are much more helpful than one), if carefully done. Let us say that if one desires to obtain the radius of a regularly pulsating star, one needs to obtain radial velocities around the same time that the optical and IR photometry are carried out. For such details and the reduction and analysis procedures, one can read any of many sources. One close at hand is that of Milone *et al.* (1999); additional references are found therein.

Eruptive objects such as novae and supernovae develop dust shells that are studied optimally in the infrared. Generally, any object with a dust shell can be studied photometrically in the IR to find the shell's composition as well as its temperature.

On the other hand, the dimming of starlight due to the scattering effects of interstellar dust is minimal in the infrared; this fact provides important clues to the intrinsic brightness and reddening at shorter wavelengths.

Thus, there are many potential and interesting infrared targets available for study, and too few astronomers to study them all. Moreover, they can be studied to a greater accuracy and precision than previously, thanks to new filter designs.

### 3. The challenges of infrared photometry

Infrared photometry is beset by difficulties, but most of these afflict mainly the thermal infrared (beginning at the K-band window of the atmosphere, roughly the region between 2 and 3 micrometers, or 20,000 to 30,000 Å). If the temperature of the telescope and observing site is 20° C, the environment is radiating most strongly at ~10 micrometers. Therefore, the telescope and photometer must be configured to minimize the effect of this local radiation. First, the optics must be pristine so that dust and other material on the mirror surfaces do not radiate onto the detector. Next, the optics should be of high quality so that only the intended radiation is focused onto the detector. If the telescope is of a Cassegrain design, for example, such that the primary mirror has a central hole to pass the radiation to the detector from a secondary mirror suspended above it on the optic axis, it is important that only the desired stellar radiation reach the detector. To accomplish this, the secondary mirror must not reflect into the detector any part of the warm primary mirror cell structure, so a stop must be used in the system if it does, to crop the radiation from outside the edge of the primary mirror itself. If a single-detector photometer is to be used to observe at wavelengths longer than 3.5 micrometers, a bolometer may be used, and this requires liquid helium as a coolant, a cryogen that is cooled to at most a few degrees above absolute zero (–273° C). The cooler the detector, the less thermal noise that it produces, and the greater its sensitivity. Usually the detector is mounted within a double-walled dewar with liquid nitrogen in an external jacket and liquid helium in an inner jacket. The outer jacket keeps the inner container and the detector cool through slow evaporation, a cooling process; it must be refilled periodically. The transfer of liquid helium is, itself, an art and the procedure can be dangerous if mistakes are made. Very warm work gloves must be used to handle transfer lines and dewars because of the extreme cold. Water vapor may freeze in the narrow neck of the dewar, blocking evaporation of the helium, and as the liquid helium warms up it will evaporate well before the water, nitrogen, and oxygen ices sublime, potentially resulting in a catastrophic explosion.

Even when the optics are pristine, and all transfer tasks are handled properly, the photometry must be done with great care for the results to be meaningful. For example, the bright warm sky background surrounding the star image must be subtracted from the stellar observation, so, even in this day of multi-detector arrays, the secondary mirror must be "chopped" (made to oscillate at some frequency between 10 and 50 Hz, typically), so that a star-free field can be subtracted from the field containing the star and the difference co-added to create a star-only image. Because the chopping is done to one side of the star, a further precaution is to "nod" the telescope (move the entire telescope) by a distance equal to the chop amplitude, so that the starless and star-containing apertures are both 180° different in phase, and the sky that is sampled is on the other side of the star along the chopping line. This periodic nodding, done at a much lower frequency than the chopping, averages and smooths-out variations of background infrared sources across the region of the target. It helps to be able to change the orientation of the chopping and nodding line so that hot spots on the sky can be avoided altogether. In this way, the constantly varying sky background can be subtracted as quickly as possible and asymmetries in the background near the target can be minimized. When an extended object such as a star cluster, or a nebula, rather than a single star, is to be observed, the precautions must be similar. Typically, the secondary mirror is under-sized in the sense that when the secondary is tilted it still does not pick up radiation from beyond the edge of the primary mirror.

In all circumstances, the filters to be used must also be as free as possible from radiation from the atmosphere; they should not, for example, include water-vapor emission. Yet, this is commonly the case with the previous, and almost all of the present, generation of infrared filters.

For the so-called "non-thermal" infrared, at wavelengths shorter than about 2.5 microns, the observing situation is less formidable, because the telescope and environment are not radiating strongly at wavelengths near the radiation to be measured, but nevertheless can be daunting. In this spectral region, the preferred coolant has been liquid nitrogen, typically, and both the inner and outer jackets are filled with this cryogen, which is certainly easier to handle than liquid helium, but of course still hazardous because of its extreme cold (~77 K or –196° C, or less). Infrared arrays do not require the chopping and nodding that is done in the thermal IR, but they do require flat-fielding, dark current, and bias treatment as for CCDs, with the added requirement that several images be "dithered" (images taken of fields differing by a few arc-seconds or more in position) and then median-combined so that a star-less image of the background can be obtained and removed from the target field. Of course, if the detector is not an array but a single detector of InSb, HgCdTe, or other substrate, the same nodding and chopping techniques as for bolometers must be used.

The infrared dewar and lock-in amplifier for single-channel photometry at the Rothney Astrophysical Observatory at the University of Calgary can be seen in Figures 2a and 2b, respectively. Cryogenic dewars such as that of Figure 2a (manufactured by Infrared Labs in Tucson) are expensive, but contain all the necessary optics for detection. In this case, it contains a single detector cooled to liquid nitrogen temperatures. The lock-in amplifier obtains a signal that is modulated at the chopping frequency of the secondary mirror (hence the "lock-in"); the signal is then rectified and integrated for a specified interval.

Even for the near IR, the telescope optics must be clean, and sources of stray light and heat minimized. The availability of a thermoelectrically cooled IR-sensitive photodiode is an important new development and deserves special attention, because LN2 dewars and the detectors within them are very expensive, even if only the *z, J,* and *H* windows can be explored (the windows are described and discussed in the next section). Henden (2002) suggests how amateur astronomers can make use of such a photometer (an Optec, Inc. SSP-4 photometer with an InGaAs PIN photodiode detector), and we echo his suggestions for careful observational procedures with this non-chopping system; we recommend, however, a change of filters for use with this photometer.

Even within these two atmospheric windows, the filters that are available have not been optimized for most observing sites to pass the extra-atmosphere radiation and exclude that from the atmosphere alone. The best of these non-optimal filters, the "Mauna Kea" set, was not produced after extensive optimization experiments in both placement and width within the windows but apparently selected to merely butt up against the atmospheric windows as modeled for the Mauna Kea Observatory, one of the highest and driest sites in the world. At less favorable sites, the atmospheric windows are narrower and the passbands will be defined by the edges of the atmospheric windows (discussed in the next section). This is not the way to achieve the best photometry! Consequently, it is not surprising that our tests have shown (Milone and Young 2005; 2007) that even at Mauna Kea itself, the Mauna Kea set do not appear to be optimal, generally, and those filters are, at best, only nominally useful at intermediate elevation sites (~2 km) such as Kitt Peak and Cerro Tololo, primarily because of the variability of the water vapor content of the atmosphere; moreover, our tests indicate that they are inadequate for the lower altitude sites (~1 km and lower) at which most university and amateur observatories are located. Therefore, we strongly advise astronomers, both amateur and professional, to seek the most suitable passbands with which to carry out their infrared photometry.

It is precisely the optimization of filters that we now wish to discuss so that optimal infrared photometry can be carried out by all astronomers, whatever their astronomical targets.

## 4. The IRWG passbands and how they improve infrared photometry

Figure 3 shows the simulated transmission of the Earth's atmosphere for a standard model atmosphere normalized to 1 and computed for an observatory site at 2.1 km elevation, typical of large multi-user observatories such as the Kitt Peak National Observatory near Tucson, Arizona. The lower abscissa scale is in wavenumbers, with equivalent micrometers on the upper scale. The spectral windows of the atmosphere, where some light is transmitted, are indicated by the designations z (for the 1 micron region), *J, H, K, L, M,* and *N* (for the 10 micron region). Except for "*z*" and "*H*," these are the designations of the original Johnson (1965; 1966) passbands which, however, were not optimally placed within these windows, and hence, provided relatively poor photometry because the boundaries and even transmission within the windows varies strongly with water vapor content of the atmosphere. Although the filters have been manufactured many times with differing central wavelengths and spectral widths, they have been designated usually the same way, according to which window they were mainly designed to cover. Consequently, it seems reasonable to refer to these "windows" with those designations.

In the late 1980s a joint commission meeting on infrared photometry was held at the General Assembly of the International Astronomical Union held in Baltimore, Maryland. The meeting explored the reasons for the problems of lack of reproducibility in infrared data and an overall limitation of about 3% in precision (from comparisons of near-infrared photometry from different observatory sites and even at the same site at different instants. See Milone (1989)). The result was the creation of an infrared working group (IRWG) of Commission 25 (Photometry and Polarimetry) which, among other tasks, was to redesign the broadband infrared photometry system by placing and shaping a new set of passbands within the atmospheric windows optimally. This it has done (Young *et al.* 1994; Milone and Young 2005, 2007). We now call the passbands corresponding to these filters *iz, iJ, iH, iK, iL, iL'* (in a slightly longer wavelength part of the *L* window), *iM,* *iN, in* (for a narrower passband in a branch of the *N* window), and *iQ*, where the "*i*" indicates "improved."

However, most amateur astronomers will most likely work in the near-infrared, where a commercial photometer, Optec, Inc., SSP-4, with a Hamamatsu photodiode is available. Therefore, we will confine our comments to the non-thermal region, which includes the *z, J,* and *H* windows. The references given above can be consulted for the improved passbands in the other windows.

A comparison of a measure of the signal to noise ratio and the atmospheric (water vapor) extinction coefficient for all of the IRWG improved filters has been made for all existing passbands and these have been compared to the IRWG simulated and manufactured filter sets. We show annotated plots of these relations in Figures 4 and 5 for the *z* and *J* windows and for the *H* window, respectively, for a particular atmospheric model atmosphere. The passbands with the highest signal-to-noise and lowest atmospheric extinction characteristics lay in the upper left parts of these diagrams. For further discussion of our S/N statistic, see Milone and Young (2005). In Figure 4, the designations "*sJ*" is for a Mauna Kea filter, and "*sJn*" is a very narrow filter in a clean part of the window devised for calibration purposes. In Figure 5, Note that the Mauna Kea filter "*sh*" lies very close to the older RAO "*H*" filter, designated "*rH*."

One of the problems with conventional filter photometry in the IR is that water vapor causes great amounts of light to be absorbed from the light contained within those filters in the beam from the star to the telescope. This causes a curvature in the extinction curve between 0 and 1 air mass (0 air mass is that at the top of the atmosphere; an air mass of 1 is the path of light impinging vertically on the observer at the observing site; 2 air masses is the path length for a zenith angle of 60°, and 3 air masses is that at an angle of about 75°). This curvature is known as the "Forbes effect" (Forbes 1842) and is not easily measurable from the ground (where we cannot see to air masses less than 1). Hence, the loss due to water vapor absorption in the atmosphere is not known precisely, but it *is* variable at all time scales! The IRWG passbands minimize the extinction and the Forbes effect; even the extinction between 1 and 3 air masses is lessened. This is the reason these passbands are improved. At the highest and driest sites, they will likely pass less infrared flux than will older filters, but, as we have argued, the tradeoff is better signal-to-noise and more reproducible results, from hour to hour, and day to day, and site to site! This is the answer to the 3% problem (a typical upper limit of agreement among different data sets) that has plagued previous generations of IR passband photometry. For more readily accessible details about the Forbes effect, consult Young *et al.* (1994). A summary is also provided in Milone and Young (2005). Details of the atmospheric modeling software, MODTRAN, can be found in Berk *et al.* (1989).

The extinction curves between 1 and 3 air masses for the *iz, iJ, iH,* and *iK* passbands for two standard stars can be seen in Figures 6 and 7. The extinction data in Figure 7 were obtained on the same date as the data obtained in older *JHK* passbands *rJ, rH, rK,* shown in Figure 8. The current practice in the literature is to place primes on the *JHK* designations to indicate narrower passbands which may or may not be repositioned within the windows so one occasionally sees a *K'* passband mentioned in the literature. In any case, none of these and none of the "*r*" filters have been optimized to the same extent as the IRWG passbands. Therefore it is not surprising that we see much larger extinction between 1 and 3 air masses with the *rJ, rH,* and *rK* filters than with the IRWG near-infrared set. What is not shown here is simulated behavior of these filters between 0 and 1 air mass. In fact, there is a strong Forbes effect (strongly curved extinction between 0 and 1 air mass) for these older passbands, and a near absence of a Forbes effect for the IRWG passbands (many simulations can be seen in Young *et al.* 1994; Milone and Young 2005, 2007). This means that the true extinction and therefore the true outside-atmosphere magnitudes are not revealed by a linear extrapolation to zero air mass in the extinction curves of the older filters. The extinction lines for the IRWG passbands, on the other hand, do permit outside-the-atmosphere magnitude determination with a linear extinction approximation. Simulations reveal that in the *iz* and *iH* passbands, there is effectively no curvature in the extinction line for a mid-summer, mid-latitude model at the elevation of the RAO (1.27 km), and even in the *iJ* and *iK* passbands it is small enough to result in only a small offset from a direct linear extrapolation to zero air mass. Note that Figures 4 and 5 indicate that infrared passbands that produce lower extinction curves tend to be those with higher signal-to-noise characteristics. It therefore follows that because the manufactured IRWG filters approximate the designed passbands well-enough that data obtained with them have lower extinction, they also have higher signal-to-noise characteristics.

## 5. A note about nomenclature of the *iz* passband

There is a need to keep nomenclature as precise and meaningful as possible, so we make the following comment about the passband notation we have been using since the early 1990s.

The proliferation of photometric systems has made any unambiguous nomenclature of passbands impossible. For example, the "*z*" window has been given a variety of names, and the letters *z* and *Z* have been applied to a variety of different wavelengths. Although wide passbands near 1 micron wavelength go back to the six-color photometry of Stebbins and Whitford in the 1940s, the first use of a narrow passband in this window seems to have been by Walker (1969), who denoted the passband by *W*; he also used a passband called *Z*, but its effective wavelength was 2.2 microns. The first association of the letter *Z* with the 1-micron passband was made by asteroid photometrists in the late 1970s (see Gradie *et al.* 1978; and Tedesco *et al.* 1982), who used the letter *Z* for it. However, the Vilnius system introduced in the 1960s already used a *Z* band near 516 nm, and many other workers have used something similar, sometimes denoted by *z* instead of *Z*. Shortly after the asteroidal *Z* band was introduced, Schneider *et al.* (1983) introduced a broad band they called *z*, extending from about 840 nm to beyond 1 micron but truncated on the IR side by the rapidly decreasing sensitivity of the detector they used; unfortunately, it is badly mutilated by the water-vapor absorptions just longward of its centroid wavelength. Passbands descended from this so-called "Gunn *griz*" system have continued to be used since then in galaxy photometry. On the other hand, Hillenbrand *et al.* (2002) used a narrow passband at 1.035 microns that resembles our *iz* band, but named their band *Y*, in conflict with the *Y* passband of the Vilnius system in use for some forty years. Clearly this is an issue that needs to be considered by Commission 25 of the IAU in the first instance.

### 6. How to obtain the filters

As we have intimated, infrared astronomers have tended to tinker with the original Johnson passbands instead of abandoning them, even though the precision and accuracy of transformations of data obtained with those filters and their improved but still not optimal passbands is sharply limited. This practice continues to the present due to both economic and perhaps sociological reasons which have little to do with the best scientific practice. Nevertheless, hope springs eternal, so we are making knowledge of the availability of an optimized set of filters to a wider body of interested observers, in the hope that the new system will be sought by many, a circumstance that, we are assured, will bring down the cost of the filters, which are prohibitively expensive when produced in small batches. The sole supplier of the IRWG filters to date has been Custom Scientific of Phoenix, Arizona, but the costs of manufacturing these filters has increased greatly over the past decade. Apparently the only way to do this economically at present is to amass a large number of orders for a batch run. Therefore, all interested parties should encourage filter manufacturers to make, and photometer manufacturers to market with their photometers, the IRWG set.

We now describe some of the specifications for the near IRWG filter set. The filter spectral profiles are triangular, but experience has shown that if these have a flat spectral region at peak transmission so that the profile is in the form of a trapezoid, this does not hurt the reproducibility very much, and for some passbands, can even improve them. In Table 1, we give the central wavelength, the width at 80% relative transmission, and half-power full widths, in Ångströms (divide by 10,000 for micro-meters). We also include here the *iK* passband, for astronomers who have access to detectors sensitive to longer wavelengths.

These are the specifications for the passbands at the operating temperature of the filters and detector. For good photometric practice, the transmission outside of these limits should be strongly blocked, to better than about one part in a million. It is always a good idea to obtain transmission traces of the filters once they are obtained to check on the transmission both within and outside the specified passband.

For further details on the background and method of experimentation that led to these passbands, see the previously mentioned references, especially Young *et al.* (1994).

## 7. Conclusions

We have discussed some of the programs to which infrared photometry can contribute as well as some of the problems with infrared photometry as currently practiced and the solutions to them; in particular, we demonstrate the advantages of an optimally designed set of infrared filters to permit improved photometric precision and accuracy. We urge astronomers who have not yet attempted infrared astronomy because of past limitations to approach the manufacturers of photometers and filters to ask for the IRWG set of filters with which improved IR photometry can be carried out. In this way, and perhaps only in this way, given sufficient care in both observational techniques and reduction and analyses, can substantial progress in IR photometry be achieved, and with it, improved accuracy and precision in all appropriate infrared investigations (i.e., ground-based, broad-band) in which these passbands are used. This is No. 75 in the *Publications of the RAO* series.

Table 1.  IRWG Near IR Passbands.

| Passband | Central Wavelength | FW 80% Transmission | HPFW |
|----------|--------------------|--------------------|------|
| *iz* | 10,320 Å | 310 Å | 730 Å |
| *iJ* | 12,400 | 290 | 790 |
| *iH* | 16,280 | 870 | 1,520 |
| *iK* | 21,960 | 990 | 1,880 |

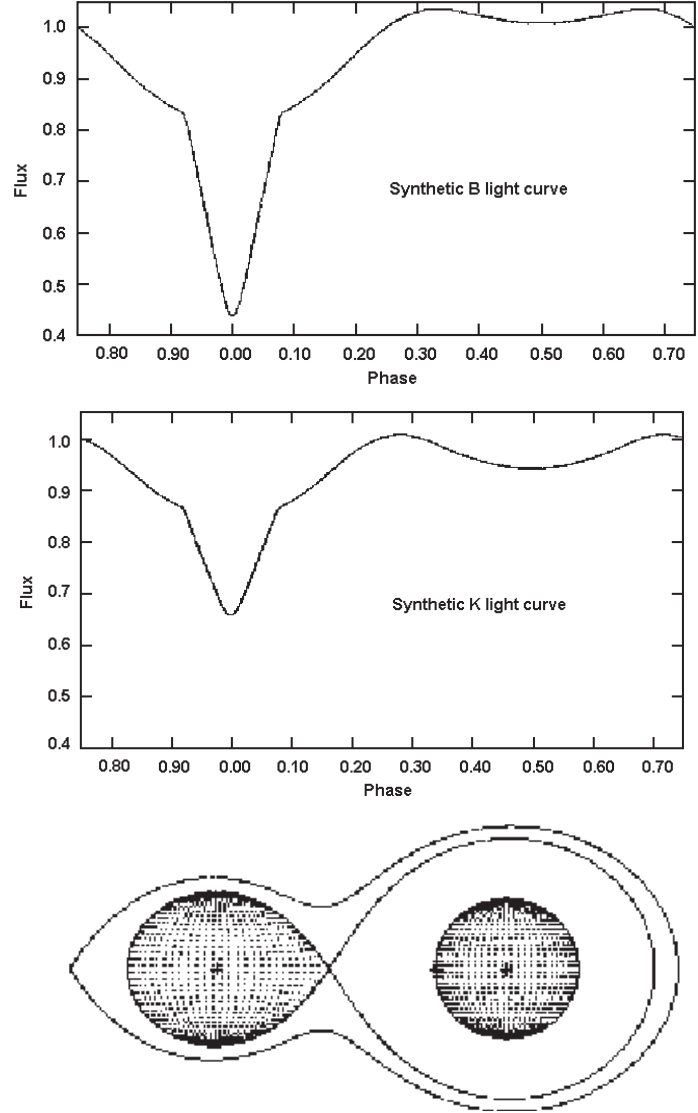

Figure 1. Top. A simulated B light curve for the Algol eclipsing binary DM Persei, from the elements of Hilditch *et al.* (1992), produced with Binary Maker software (Bradstreet and Steelman 2004), as were the lower panels. Middle. A simulated *K* light curve for DM Persei. Bottom. A Lagrangian surface plot for the semi-detached eclipsing system DM Persei, again from the elements of Hilditch *et al.* (1992). For a later model that incorporates third light from a third component, see Van Hamme and Wilson (2007).

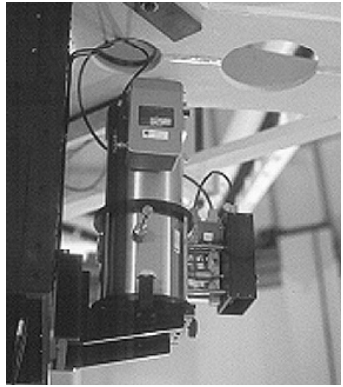

Figure 2a. An infrared dewar, with filters enclosed and kept at cryogenic temperatures, attached to a side port of the Cassegrain focus of the Rothney Astrophysical Observatory's 1.8-m Alexander R. Cross Telescope (ARCT). The light was conveyed from the telescope optics through a right-angle front-silvered diagonal mirror. Such a double dewar is used conventionally to keep the noise of the detector to a minimum.

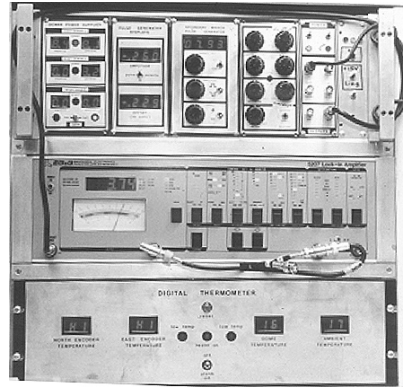

Figure 2b. A bank of electronics in the control room of the ARCT that controls the chopping of the secondary mirror (top panel) and detection of infrared radiation synchronized to it, through a Lock-in Amplifier (middle panel). The lowest panel displays temperatures at this alt-alt telescope's motor-encoders.

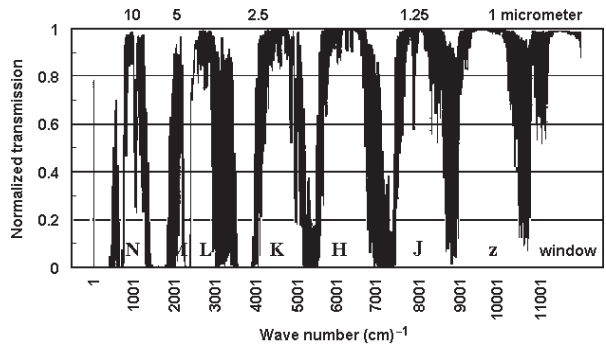

Figure 3. Infrared atmospheric transmission computed for a U.S. Standard Atmosphere model for a site at 2.1 km elevation, appropriate for such sites as the Kitt Peak National Observatory in Arizona. The wavelength in micro-meters corresponding to some of the wavenumbers on the abscissa is indicated and the atmospheric "windows" where the transmission is highest, are marked with letters. These should not be confused with passband or filter names, which often have the same designations, but often are not confined to the individual windows.

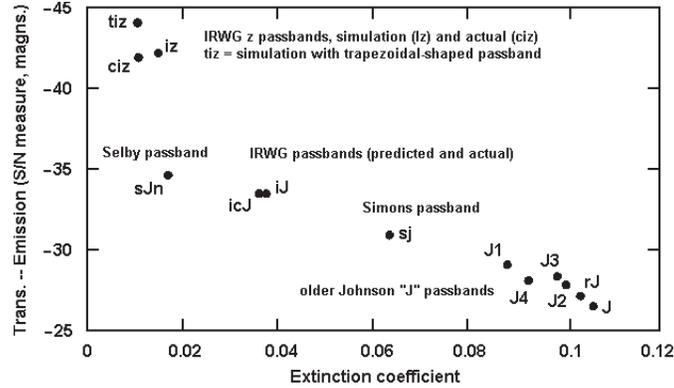

Figure 4. A plot of quality of a number of passbands for use in the *z* and *J* windows, in current use by astronomers, in the form of a computation of a measure of the signal-to-noise vs. the computed extinction coefficient. The original Johnson passband is marked with a "J" and others similar to it, have a numbered suffix, except for "rJ," which is an older passband used at the RAO. The "Simons passband" indicates the Mauna Kea passband as sent to us for testing by D. Simons in 1997. The *iJ* and *icJ* passbands indicate the IRWG passband as designed and manufactured, respectively. The *z*-window passbands of the IRWG set are similarly marked. The *tiz* designation is for a flat-topped, trapezoidally shaped passband instead of the triangular shaped for all the other IRWG passbands. The highest signal-to-noise values are at the top of the diagram. Note that there is a tendency for the passbands with the highest signal-to-noise ratio to be associated with the smallest extinction coefficients. This alone shows that the newer passbands are less sensitive to atmospheric extinction.

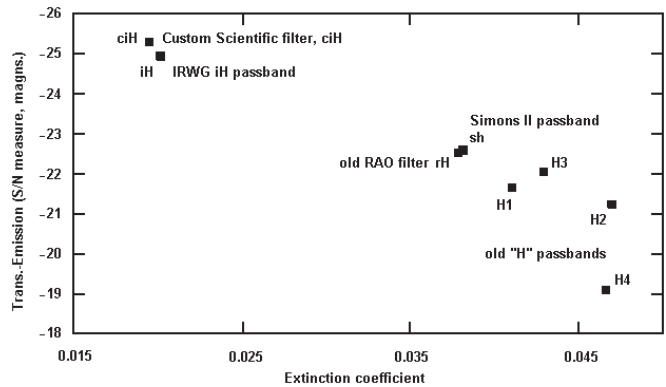

Figure 5. A quality plot for a number of passbands used for photometry in the *H* window of the atmosphere. The passbands are denoted as for Figure 4. Again, the IRWG passbands show the highest signal-to-noise ratio and the lowest atmospheric extinction. The model atmosphere used for the tests that produced the results of Figures 4 and 5 is for a summer, mid-latitude site at 1 km elevation.

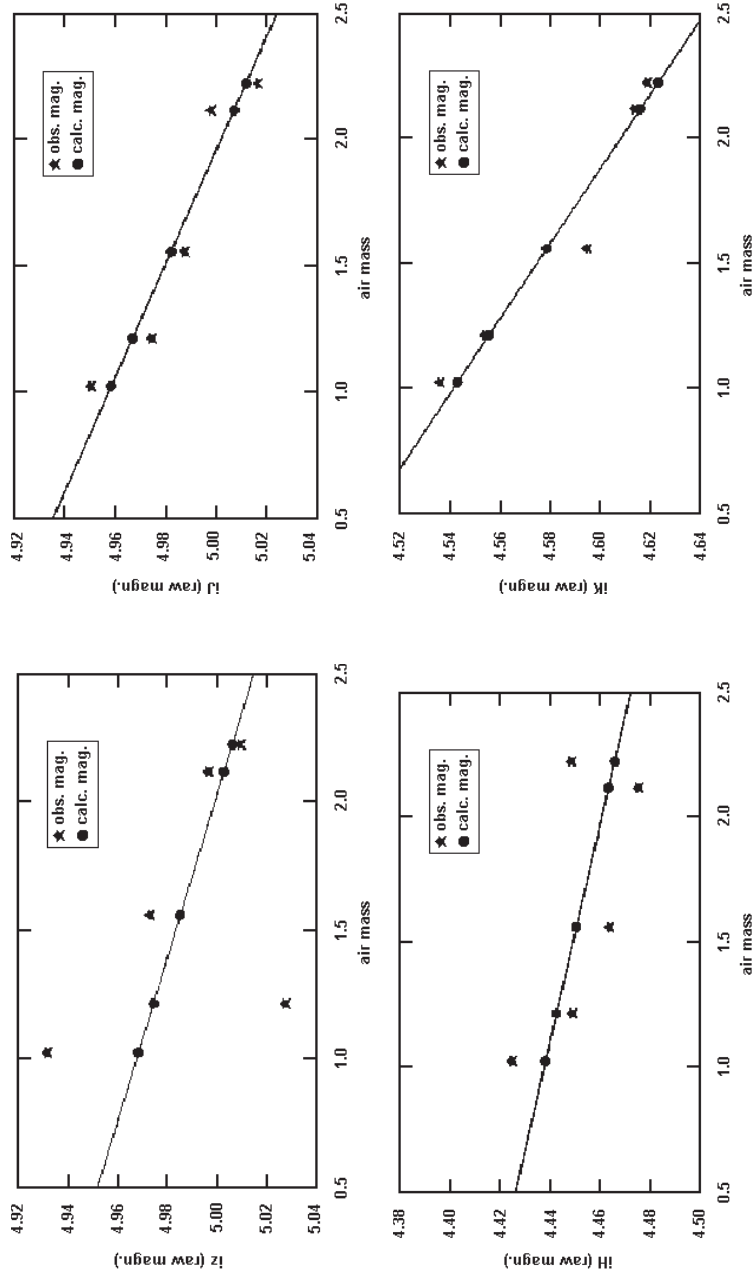

Figure 6. Observed extinction plots for the night of July 13, 2000, for the IRWG passbands, made on the RAO's 1.8-m ARCT. Linear fitting to the data may be extrapolated to give a close approximation to outside-the-atmosphere magnitudes. The extinction star was χ Herculis; k' = –0.03, 0.04, 0.02, and 0.03 magn./airmass for *iz*, *iJ*, *iH*, and *iK*, respectively.

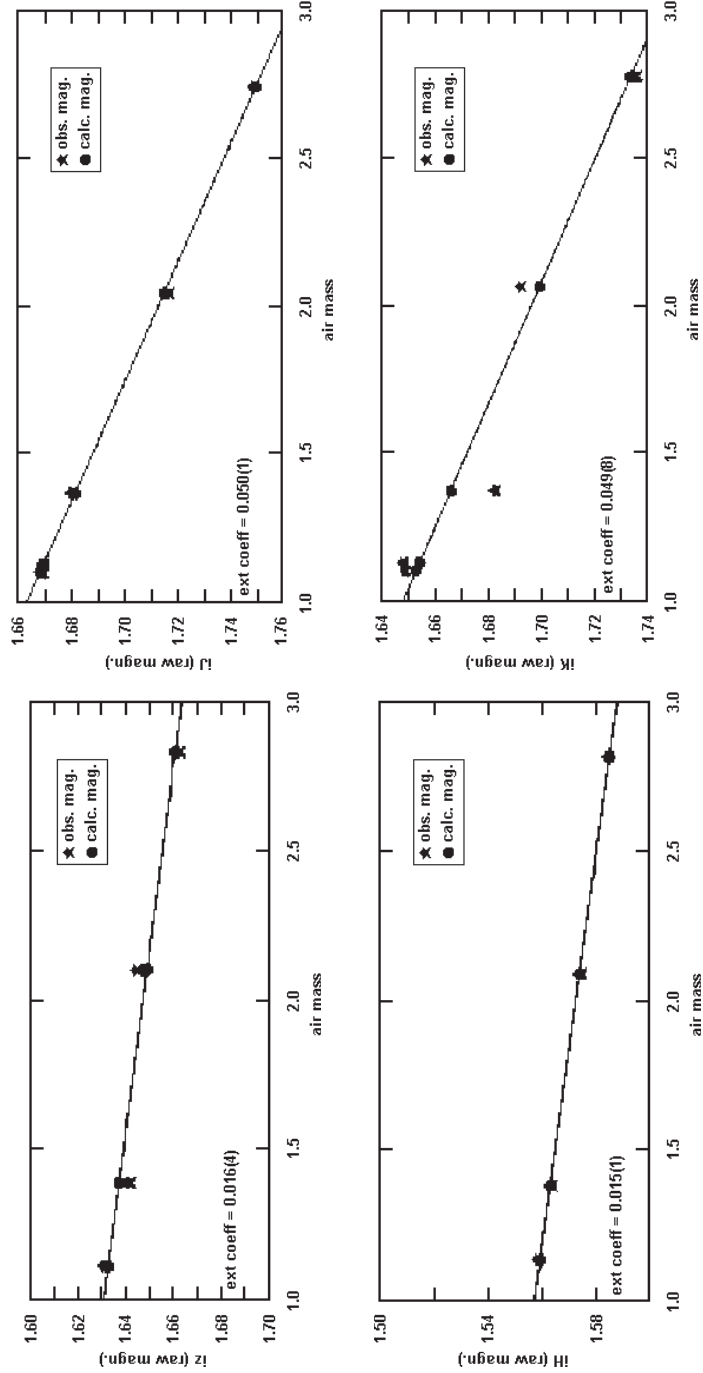

Figure 7. Observed extinction plots for the night of Sept. 26, 2000, for the IRWG passbands, made on the RAO's 1.8-m ARCT. Linear fitting to the data may be extrapolated to give a close approximation to outside-the-atmosphere magnitudes. The extinction star was Vega. The derived linear extinction coefficients and their uncertainties, in units of the last decimal place (in parentheses) are given in the lower left corners of the plots, viz., about 0.02, 0.05, 0.02, and 0.05 magn./airmass for the $iz$, $iJ$, $iH$, and $iK$ passbands, respectively.

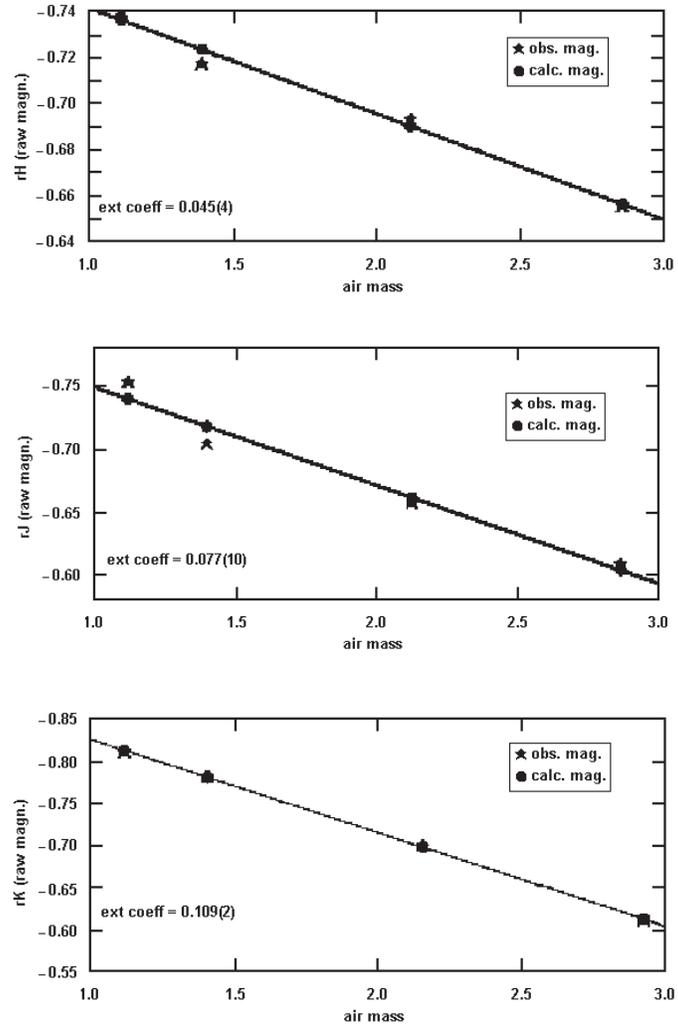

Figure 8. Observed extinction plots for the same night in which the IRWG passbands were used, and for the same extinction star, Vega, but obtained with an older set of passbands. The derived linear extinction coefficients and their uncertainties are again indicated. Note that they are systematically higher than for the IRWG passbands, about 0.05, 0.08, and 0.11 magn./airmass for the *rH, rJ,* and *rK* passbands, respectively.